\documentclass[10pt,twocolumn,superscriptaddress,showpacs,preprintnumbers,amsmath,amssymb,prb,aps]{revtex4-1}
\usepackage{graphicx}
\usepackage{pgf}
\usepackage{ulem}
\usepackage{color}

\begin{document}


\title{Direct evidence of superconductivity and determination of the superfluid density in buried ultrathin FeSe grown on SrTiO$_3$}

\author{P. K. Biswas}
\email[]{Pabitra.Biswas@stfc.ac.uk}
\altaffiliation[Current address: ]{ISIS Pulsed Neutron and Muon Source, STFC Rutherford Appleton Laboratory, Harwell Campus, Didcot, Oxfordshire, OX11 0QX, UK}
\affiliation{Laboratory for Muon Spin Spectroscopy, Paul Scherrer Institut, CH-5232 Villigen PSI, Switzerland}

\author{Z. Salman}
\affiliation{Laboratory for Muon Spin Spectroscopy, Paul Scherrer Institut, CH-5232 Villigen PSI, Switzerland}

\author{Q. Song}
\affiliation{State Key Laboratory of Surface Physics, Department of Physics, Fudan University, Shanghai 200433, China}

\author{R. Peng}
\affiliation{State Key Laboratory of Surface Physics, Department of Physics, Fudan University, Shanghai 200433, China}

\author{J. Zhang}
\affiliation{State Key Laboratory of Surface Physics, Department of Physics, Fudan University, Shanghai 200433, China}

\author{L. Shu}
\affiliation{State Key Laboratory of Surface Physics, Department of Physics, Fudan University, Shanghai 200433, China}

\author{D. L. Feng}
\affiliation{State Key Laboratory of Surface Physics, Department of Physics, Fudan University, Shanghai 200433, China}

\author{T.~Prokscha}
\affiliation{Laboratory for Muon Spin Spectroscopy, Paul Scherrer Institut, CH-5232 Villigen PSI, Switzerland}

\author{E.~Morenzoni}
\email[]{elvezio.morenzoni@psi.ch}
\affiliation{Laboratory for Muon Spin Spectroscopy, Paul Scherrer Institut, CH-5232 Villigen PSI, Switzerland}

\date{\today}

\begin{abstract}
Bulk FeSe is superconducting with a critical temperature $T_c$ $\cong$ 8 K and SrTiO$_3$
is insulating in nature, yet high-temperature superconductivity has been reported at the interface between a single-layer FeSe
and SrTiO$_3$. Angle resolved photoemission
spectroscopy and scanning tunneling microscopy measurements observe a gap opening at the Fermi surface below $\approx$ 60
K. Elucidating the microscopic properties and understanding the pairing mechanism of single-layer FeSe is of utmost
importance as it is a basic building block of iron-based superconductors.
Here, we use the low-energy muon spin rotation/relaxation technique (LE-$\mu$SR) to detect and quantify
the supercarrier density and determine the gap symmetry in FeSe grown on SrTiO$_3$
(100).
Measurements in applied field show a temperature dependent broadening of the field distribution below $\sim$ 60 K, reflecting the superconducting transition and
formation of a vortex state.
Zero field measurements rule out the presence of magnetism of static or fluctuating origin.
From the inhomogeneous field distribution, we
determine an effective sheet supercarrier density $n_s^{2D} \simeq 6 \times 10^{14}$ cm$^{-2}$ at
$T \rightarrow 0$ K, which is a factor of 4 larger than expected from ARPES measurements of the excess electron count per Fe of 1 monolayer (ML) FeSe. The temperature dependence of the superfluid
density $n_s(T)$ can be well described down to $\sim$ 10 K by simple $s$-wave BCS, indicating a rather clean superconducting phase with a gap of
10.2(1.1) meV. The result is a  clear indication of the gradual formation of a two dimensional vortex lattice existing over the entire large FeSe/STO interface
and provides unambiguous evidence for robust superconductivity below 60 K in ultrathin FeSe.
\end{abstract}

\maketitle

\section{Introduction}

Following the discovery of high-$T_c$
cuprates~\cite{Bednorz,Schilling} a few decades ago, the Fe-based superconductors~\cite{Hosono,Hsu,Johnston,Paglione,Stewart,Wang} represented an additional novel and important
class of high-$T_c$ superconductors displaying, however, average critical temperatures lower than the cuprates. Surprisingly, high-temperature
superconductivity with a $T_c$ $\approx$ 60-70 K was found in single-layer FeSe on SrTiO$_3$ (STO) ~\cite{Wang2,Lee,Liu,He,Tan}.
Similar high temperatures exceeding that of all known bulk iron-based superconductors have also been achieved on other oxide substrates \cite{Rebec2017}.

This finding is extremely important in view of the simple crystal structure of the system, which consists of a single Se-Fe-Se unit, i.e. the basic building block of all iron-chalcogenide superconductors, and may pave the way to identifying key ingredients of
high-$T_c$ superconductivity ~\cite{Lee}. Single-layer FeSe exhibits a distinct electronic structure with only electron
pockets near the Brillouin zone corner~\cite{Liu,He,Tan}. This is in contrast to its bulk counterpart, which also shows hole pockets at
the zone center.

Transport measurements performed \textit{ex situ} find, with respect to bulk, an enhancement of $T_c$ with onset around 40 K not only in 1 ML FeSe \cite{Zhang2} but also in ultrathin layers in various configurations \cite{Wang2017} including electric-double-layer transistor films \cite{Shiogai} and ultrathin flakes on SiO$_2$/Si \cite{Lei2016}. Similar $T_c$ as on STO have been measured on other substrate materials such as MgO, KTaO$_3$ \cite{Shiogai2017}, TiO$_2$ (rutile \cite{Rebec2017} and anatase phase \cite{Ding2016}) and K-doped FeSe films \cite{Wen2016}, whereas \textit{in situ} zero-resistivity was detected at a temperature as high as 109 K~\cite{Ge}. Diamagnetic shielding was also observed up to $T_{onset} \sim 65$~K \cite{Zhang2015}.



The superconducting gap of FeSe/STO has been mainly characterized by surface sensitive techniques such as ARPES and STM. The data
suggest that single-layer FeSe has plain
\textit{s}-wave pairing symmetry~\cite{Fan2015, Lee, Liu, Tan}. However on its own, detection of a gap appearing below $\sim$ 60 K does not provide conclusive evidence that
it is only related to the formation of a condensate of Cooper pairs and does not exclude other contributions such as
magnetic, charge or spin density wave gaps. Transport measurements, on the other hand, cannot easily discriminate between filamentary and bulk superconductivity.
It is therefore essential to characterize the presence of superconductivity in FeSe/STO and its microscopic properties by other techniques, providing complementary information such as
the superfluid density and the homogeneity of the superconducting phase.

Here, we report detailed depth-resolved investigation of the superconducting and magnetic properties in ultra-thin FeSe by
the low-energy muon spin rotation/relaxation (LE-$\mu$SR) technique. Zero field (ZF) $\mu$SR measurements demonstrate that the ground
state is non-magnetic and transverse field (TF) $\mu$SR results show that superconductivity appears below 62 K. Taking
into account the extreme 2D-character of the vortex state, we estimate the effective superfluid sheet density $n_{s}^{2D}(T)$. Its temperature dependence is well described down to $\sim$ 10 K by a simple BCS $s$-wave model, with a gap $\Delta(0)=10.2(1.1)$ meV.

\section{Experimental details}

\subsection{Film growth and characterization}

\begin{figure}[h]
\begin{center}
\includegraphics[width=0.8\columnwidth]{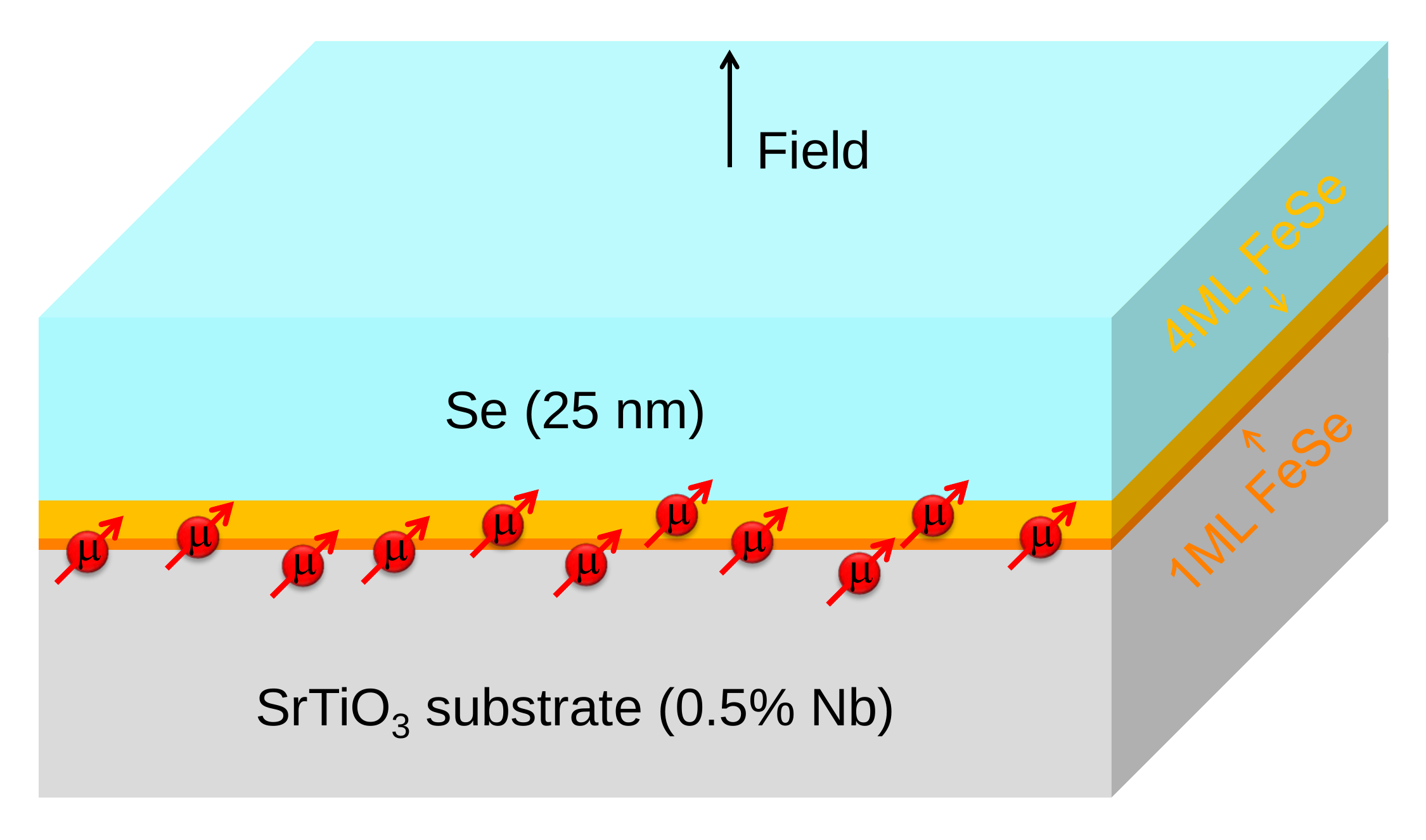}
\caption{ \textbf{Layers of the heterostructure.} Schematic diagram (not to scale) of the heterostructure with a
  ultrathin FeSe film grown on the SrTiO$_3$ substrate. For transverse field measurements the magnetic field is applied perpendicular to the sample surface.
  The polarization of the implanted muons is parallel to the sample surface.
}
 \label{fig:heterostructure}
\end{center}
\end{figure}

\begin{figure}[h]
\begin{center}
\includegraphics[width=1.0\columnwidth]{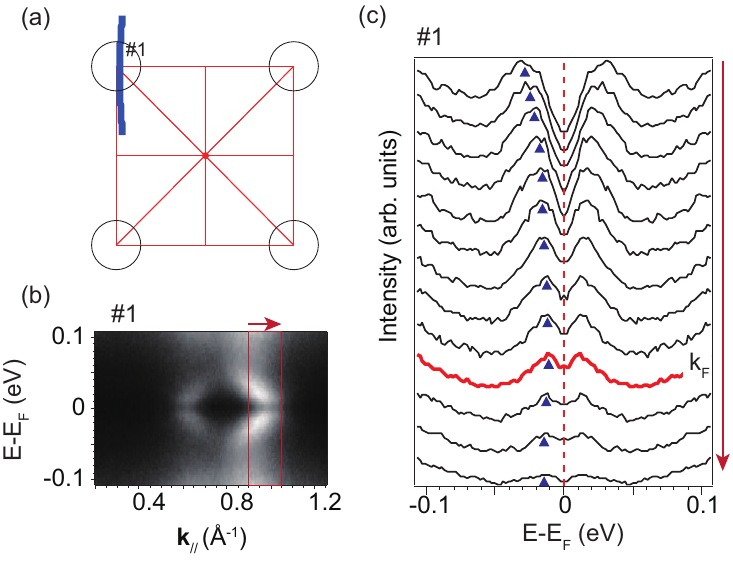}
\caption{\label{ARPES_EDC} \textbf{ARPES measurements} \textbf{(a)} Sketch of the Fermi surface sheets and Brillouin zone of single-layer FeSe/STO. Cut \#1 is indicated in the Brillouin zone. \textbf{(b)} The photoemission intensity along cut \#1, which is symmetrized with respect to Fermi energy. \textbf{(c)} Symmetrized energy distribution curves along a portion of cut \#1 which is indicated by the red arrow in panel b. Data were collected at 25K.}
\end{center}
\end{figure}

Figure~\ref{fig:heterostructure} shows a schematic of the heterostructure used in this experiment.
Single-layers FeSe thin films were
grown using molecular beam epitaxy (MBE) on a 10 $\times$ 10 mm$^2$ TiO$_2$ terminated and Nb-doped (0.5\% wt)
(001)-oriented SrTiO$_3$ substrate.
%
The substrate was pre-cleaned
following the method described in previous work~\cite{Tan} and ultrahigh vacuum (UHV) condition was maintained  during deposition to
enable continuous in situ growth.
In the UHV chamber the substrate was degassed at 550 C$^\circ$ for three
hours and then heated to 950$^\circ$C under a Se (99.9999\%) flux for 30
minutes. It was kept at 490 C$^\circ$ in Se and Fe
(99.995\%) flux for co-evaporation and co-deposition with the flux
ratio of ~20:1.
After growth, the films were annealed at
600$^\circ$ in vacuum for 3h. In-situ measurements confirmed the possible ~60 K
superconductivity in the monolayer (ML) FeSe film.
Four more unit cells of FeSe thin films were successfully grown
above the single-layer FeSe. The additional layers were deposited for stabilization purpose, since, surprisingly
the original tunneling spectra of two unit cells or thicker FeSe films did not show
signs of superconductivity \cite{Wang2}.
Before depositing the overlayers the FeSe ML was characterized by ARPES.
Figure \ref{ARPES_EDC} shows the result exhibiting the typical features of the electronic structure \cite{Tan,Liu}.
ARPES measurements indicated charge transfer from the substrate and superconductivity to be restricted to the FeSe interface layer with the top layer displaying charge neutrality. However, in our discussion below we will also address the question of the possible contribution of these additional layers to the observed supercarrier density, in view of our and recent results of charge distributions in ultrathin films \cite{Shiogai2017}.
A $\sim$ 25 nm thick layer of amorphous Se was added for protection. Thickness of
the films was monitored using a crystal oscillator and
confirmed by X-ray reflectivity measurements.
A susceptibility measurement by mutual induction on a sample of similar composition and structure, grown under the same condition and equipment as the sample presented here,
provided unambiguous evidence for the onset of Meissner effect at 65 K \cite{Zhang2015}.
The $\mu$SR measurements reported here were performed on a mosaic of 3 pieces of the 10 $\times$ 10 mm$^2$ surface area films.
The samples were glued to a Ni coated Al plate and mounted onto a cold finger cryostat. Ni suppresses the $\mu$SR signal from the muons not hitting the sample \cite{Saadaoui}.

\subsection{Low-energy $\mu$SR}

To measure the local magnetic and superconducting properties of the ultrathin FeSe layer we
use LE-$\mu$SR as a sensitive magnetic probe \cite{Morenzoni2004}.
Fully polarized muons are implanted in the sample one at a time, where they thermalize
and act as sensitive magnetic microprobe. The muon spin precesses around the local magnetic field
$B$ at the muon site with the Larmor frequency $\omega_{\mu}=\gamma_{\mu} B$,
$\frac{\gamma_{\mu}}{2\pi}$=135.5 MHz/T.
The precession and relaxation of the spin ensemble
leads to a temporal evolution of the polarization, which is easily
detectable via the asymmetric muon decay (lifetime $\tau_{\mu}$=2.2 $\mu$s), where a positron is emitted
preferentially in the direction of the muon spin at the moment of the
decay. From the damped precession signal the field
distribution associated with the vortex state can be
determined.
The LE-$\mu$SR experiments were performed on the LEM instrument, at the $\mu$E4 beamline of
the Paul Scherrer Institut in Villigen, Switzerland \cite{Prokscha}.
Here the energy of the muons can be tuned ($\sim$ 1 to 30 keV) to control the implantation
depth in the range ($\sim$ 1-300) nm and thus to probe the magnetic response in different layers of
the heterostructure \cite{Morenzoni}. With this unique ability, the LE-$\mu$SR technique is an ideal probe for
studying the superconducting properties of the FeSe layer by
implanting the muons on or very close to this layer. This procedure
has been successfully applied to address related questions in a
variety of systems and heterostructures. In particular, by varying the
implantation energy of the muons, the spatial evolution of the magnetic
field distribution as the flux lines emerge through the surface of a
superconducting YBa$_2$Cu$_3$O$_{7-\delta}$ film has been monitored
\cite{Niedermayer}, superconducting proximity effects of buried
cuprate layers  \cite{Morenzoni11}, the paramagnetic Meissner effect due to spin triplet components \cite{DiBernardo}
and magnetism at transition metal-molecular interfaces have been detected \cite{Al Ma’Mari}.

\subsection{Zero-field and transverse-field $\mu$SR measurements}

\begin{figure}[h]
\begin{center}
\includegraphics[width=1.05\columnwidth]{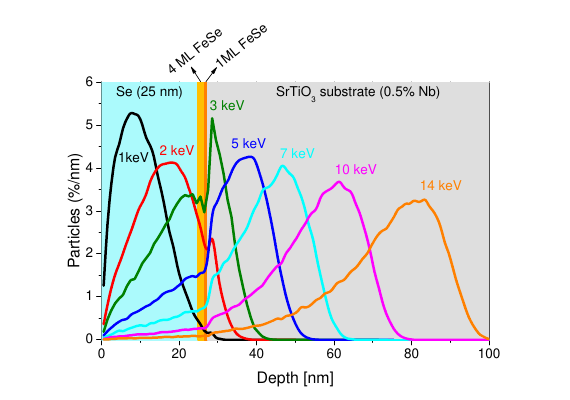}
\caption{ \textbf{Muon implantation profiles.} Muon stopping profiles
  in the investigated heterostructure calculated at different implantation energies using
  the Monte Carlo code TRIM.SP modified for muon implantation.}
 \label{fig:implantation_profile}
\end{center}
\end{figure}

Initially, we tuned the muon beam implantation energy $E$ to maximize the fraction of muons stopping in the vicinity of the FeSe single-layer. Monte Carlo simulations, presented in Figure~\ref{fig:implantation_profile}, show that this is achieved for $E \sim 3$~keV.
The program TRIM.SP, specially modified for muon implantation in heterostructures and whose
reliability to calculate stopping profiles has been previously tested, was used for the calculation \cite{Eckstein,Morenzoni2002}.

We performed ZF and TF-$\mu$SR measurements at different temperatures.
A ZF measurement is very sensitive to magnetism;
in a magnetic environment, well defined precession frequencies may be
observed in the case of long-range order. Alternatively a distribution of
precession frequencies with the corresponding width proportional to
the field inhomogeneity may be detected. If the field distribution is broad when
averaged over the sample, as in the case of disordered or short range
magnetism, the muon decay asymmetry displays a fast depolarization. In the
case of dynamic moments with fluctuating times within the $\mu$SR time
window, spin relaxation is also observed. These features allow the direct
observation of the onset of magnetic order even if very weak. It has been used for
instance to search for time-reversal symmetry breaking phenomena
in the superconducting phase, where a very tiny spontaneous
static magnetic field appears with the onset of superconductivity \cite{Luke2,Biswas}.
\begin{figure}[h]
\begin{center}
\includegraphics[width=1.0\columnwidth]{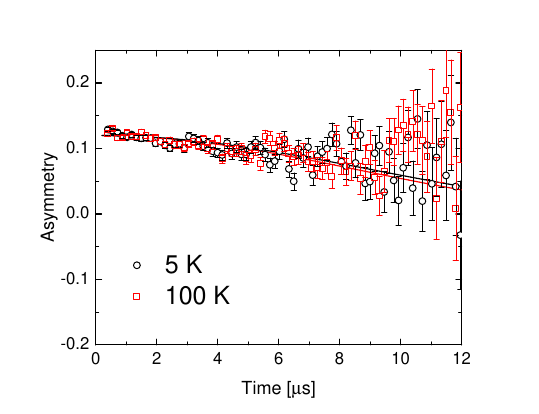}
\caption{\label{ZF-asymmetry} \textbf{ZF muon spin relaxation.}
  ZF-$\mu$SR time spectra collected at 5~K and 100~K for single-layer
  FeSe with muon implanted at an energy of 2.3 keV. The solid lines are fits to the
  data. See Supplementary Information for details about the fit function.}
\end{center}
\end{figure}
The ZF-spectra taken at 2.3 keV muon implantation energy can be described well using a static Gaussian Kubo-Toyabe
relaxation function~\cite{Kubo}, where the time evolution of the
asymmetry $A(t)$, which is proportional to the muon spin polarization,
is given by:
\begin{equation}
A(t)= A_0\left\{\frac{1}{3}+\frac{2}{3}\left(1-{\sigma^2_{\rm ZF}}t^2\right){\rm exp}\left(-\frac{{\sigma^2_{\rm ZF}}t^2}{2}\right)\right\},
 \label{eq:KT_ZFequation}
\end{equation}
where $A_0$ is the initial asymmetry and $\sigma_{\rm ZF}$ the muon
spin relaxation rate. We do not detect any difference in the spectra, taken at 5~K and 100~K, as shown in
Figure ~\ref{ZF-asymmetry}. The nearly equal and very small values of $\sigma_{\rm ZF}$ (0.086(5)
and 0.082(5) $\mu$s$^{-1}$ for 5 and 100 K, respectively), extracted
from the fits for two different temperatures, reflect the presence of
random local magnetic fields arising solely from the nuclear moments in the sample.

For the TF-$\mu$SR measurements as a
function of temperature, the sample was cooled in a magnetic field of 10 mT applied
normal to the sample surface and to the initial muon spin
direction.

\begin{figure}[h]
\begin{center}
\includegraphics[width=1.0\columnwidth]{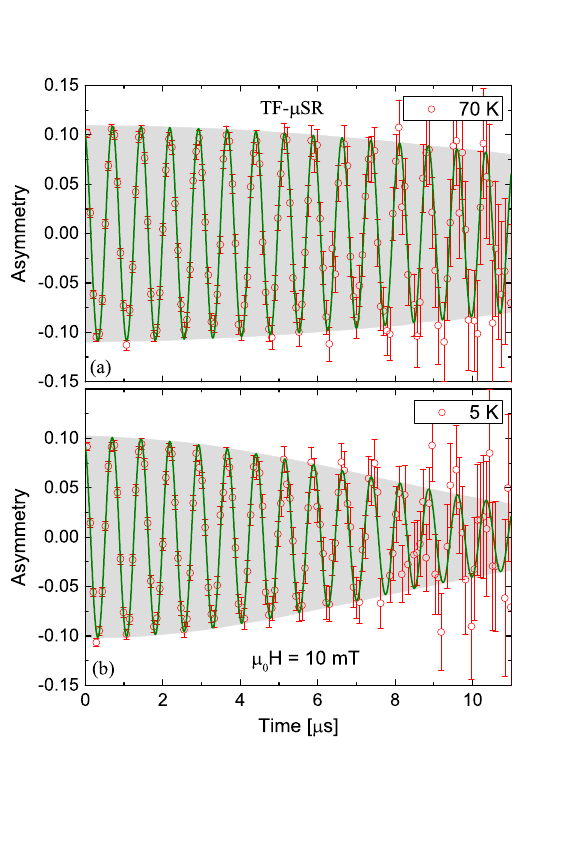}
\caption{\label{asymmetry_TF} \textbf{Muon spin rotation signal.}
  TF-$\mu$SR time spectra collected in a
  transverse field of 10~mT with a muon implantation energy of 3 keV at temperatures
  (a) 5~K and (b) 70~K . The solid lines are fits to the data
  using the Eq. ~\ref{Depolarization_Fit}. The
  shaded area evidences the different damping rate.}
\end{center}
\end{figure}

Figure~\ref{asymmetry_TF}
shows the TF-$\mu$SR time spectra collected at (a) 5~K and (b)
70~K. At 70~K, the local field probed by the muons corresponds to the applied field
and only a weak damping of the signal is observable,
consistent with the ZF results at 100 K. By contrast, the data collected at 5~K shows a more
pronounced damping. The $\mu$SR time spectra (Figure ~\ref{asymmetry_TF}) were analyzed using a Gaussian damped spin
precession signal \cite{Suter}:
\begin{equation}
\label{Depolarization_Fit}
A(t)=A_0\exp\left(-\sigma^{2}t^{2}\right/2)\cos\left(\gamma_\mu B t +\phi\right),
\end{equation}
where $A(0)$ is the initial asymmetry, $B$ is the
magnetic field at the muon sites, $\phi$ is the initial phase
of the muon polarization precession signal, and  $\sigma(T)$ is the spin damping rate due to the field inhomogeneities.

\section{Results and discussion}

\subsection{Temperature and energy dependence of the field broadening}

\begin{figure}[h]
\begin{center}\includegraphics[width=1.0\columnwidth]{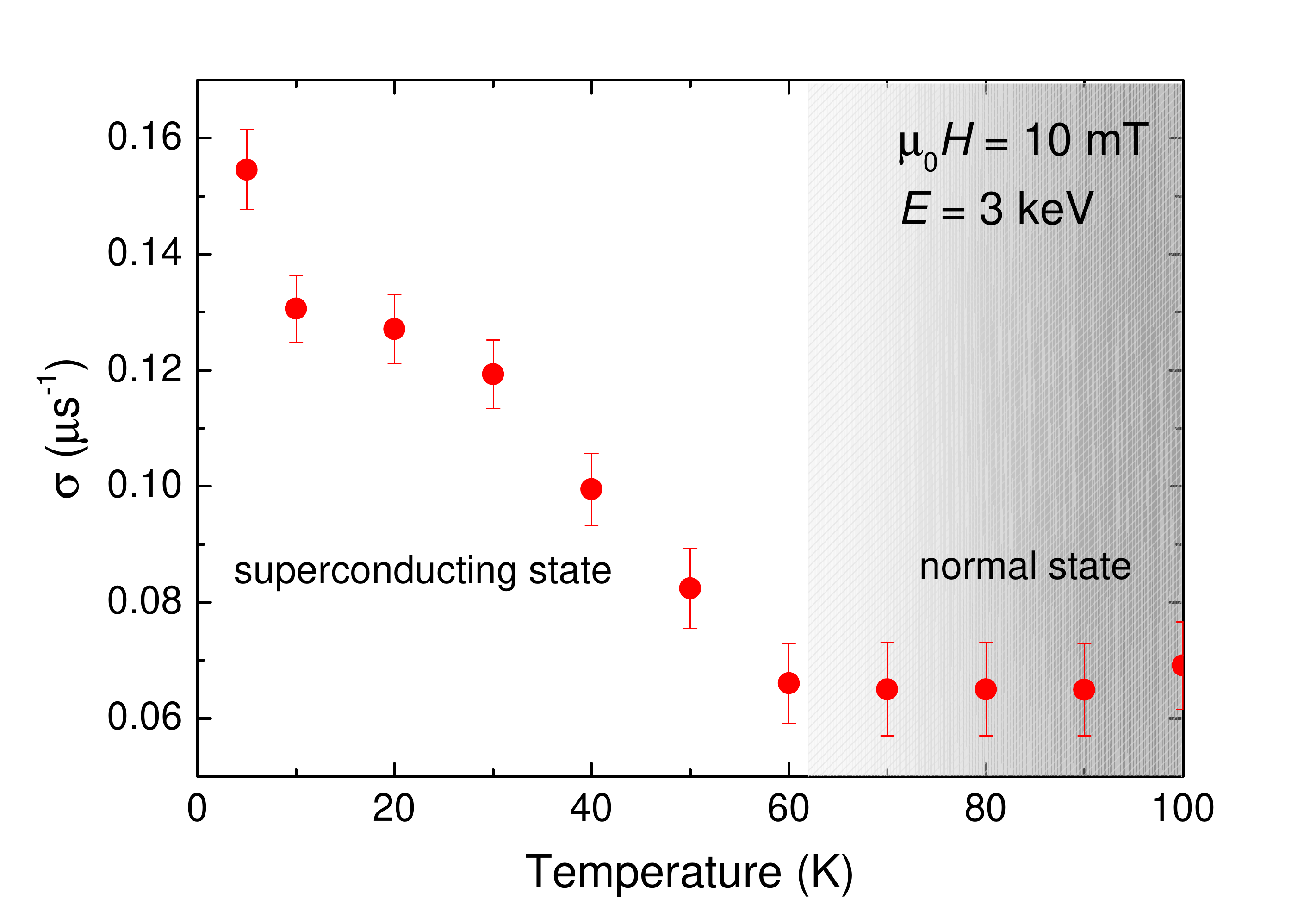}
  \caption{\label{sigma_temp} \textbf{Temperature dependence of the
      muon spin damping rate.} The temperature dependence of the muon spin damping rate $\sigma$
    measured at an implantation energy of 3 keV and an applied field of 10 mT.}
\end{center}
\end{figure}

The temperature dependence of the Gaussian damping rate $\sigma(T)=\left(\sigma^{2}_{\rm sc}(T) +
\sigma^{2}_{\rm nm}\right)^{\frac{1}{2}}$ is shown in Figure~\ref{sigma_temp}. The data displays a clear
increase of $\sigma$ with lowering the temperature due to the term $\sigma_{\rm sc}(T)=\gamma_{\mu}\sqrt{\Delta B^2}$, which expresses
the inhomogeneous field distribution associated with the formation of the vortex state in superconducting FeSe below $\sim 60$~K.
$\sigma_{\rm nm}$ ($\approx \sigma_{\rm ZF}$) is caused by the dipolar field
contribution of the nuclear moments and is temperature independent.
The average spin precession frequency, which is proportional to the average local field,
corresponds very closely to the applied field as expected from a demagnetizing factor close to one
in our geometry.
Our \textit{ex situ} value of $T_c$ agrees well with the temperature for gap opening observed in several \textit{in situ} ARPES measurements ~\cite{Wang2,Lee,Liu,He,Tan}.

The measurement at 5 K of $\sigma$ as a function of depth by varying the muon
implantation energy, $E$, further establishes the source of the observed superconductivity.
\begin{figure}[h]
\begin{center}
\includegraphics[width=1.0\columnwidth]{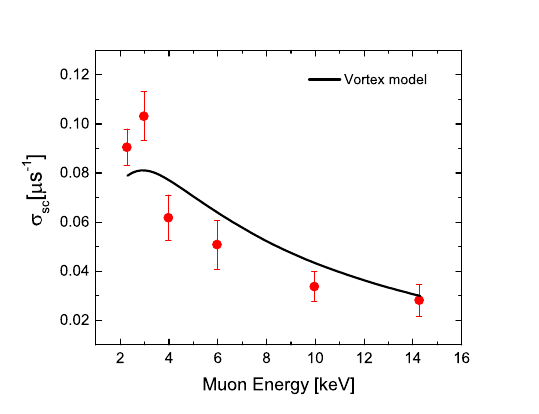}
\caption{\label{sigma_energy} \textbf{Energy dependence of the field broadening.} Muon spin
  damping rate $\sigma_{\rm sc}$ after correction of the nuclear moments contribution plotted as a function of muon implantation energy
  $E$, 5 K. The solid line shows the fit with the $\sigma_{\rm sc}$ vs. $E$ curve calculated within the London model of a very thin superconducting layer as described in the text.}
\end{center}
\end{figure}
As expected from the TRIM.SP calculations, we observe the largest field inhomogeneity at $\sim$ 3 keV, where most of the muons are implanted very close to the FeSe layers. $\sigma_{\rm nm}$ is small and temperature independent but slightly depends on the muon
implantation energy due to the different nuclear moment contribution in the various layers composing the heterostructure. We determined this contribution by performing a full energy scan in the normal state at $T=100$K
and corrected for it to obtain the energy dependence of the field broadening $\sigma_{sc}$ in the vortex state of FeSe (Fig.~\ref{sigma_energy}).


\subsection{Calculation of the field width}

$\mu$SR has been widely used to characterize the properties of bulk superconductors and determine their microscopic parameters \cite{Sonier}.
For a bulk superconductor in the vortex state the field broadening is directly given
by the magnetic penetration depth $\sigma_{\rm sc} \propto \frac{1}{\lambda^2}$. In our sample
$\sigma_{\rm sc}(T)$ is determined by
the 2D pancake-like vortices that form in a thin superconducting layer
\cite{Clem,Brandt2005a}. Since the muon stopping profile encompasses a region
outside the single FeSe layer (see Figure \ref{fig:implantation_profile}), the inhomogeneous stray field of the
vortices, which extends outside the superconducting layer
\cite{Brandt2005a, Carneiro2000}, has to be taken into account to obtain the
relationship between $\sigma_{\rm sc}(T)$ and the effective superfluid density in FeSe.

The field profile and distribution have been obtained by solving the London equation,
which is appropriate for an extreme type-II superconductor ($\xi<<\lambda$, $\xi$ coherence length $\sim$ 2-3 nm \cite{Shiogai}).
For the ultrathin FeSe layers application of a magnetic field will lead to the formation of a
regular vortex structure of hexagonal symmetry,
with each vortex carrying a flux quantum $\Phi_0$ and intervortex separation $D \equiv
\sqrt{2\frac{\Phi_0}{\sqrt{3}B_0}} \cong$ 490 nm for $B_0=10$ mT.
Indication of such a structure has been visualized by STM measurements \cite{Fan2015}.
In a bulk superconductor the local field $B_z(x,y,z)$, although varying with the planar coordinates $x$
and $y$, is always parallel to the applied field and perpendicular to the sample surface ($z$ direction, $z$=0 center of the single layer).
In our
case, near the single-layer, the field lines splay out. However, this
effect on the $\mu$SR signal is small and we can consider the normal component of the field \cite{Brandt2005}.

We determine $B_z(x,y,z)$ from the requirement that it fulfils London equation with source terms representing the flux lines core in a very thin superconducting film ($-d/2<z<d/2$)
and Laplace equation outside
\begin{multline}
-\nabla^2 B_z(x,y,z) + \Pi(z)\frac{B_z(x,y,z)}{\lambda^2}= \Pi(z)\frac{\Phi_0}{\lambda^2}\sum_{\vec{R}} \delta(\vec{r}-\vec{R})
 \label{eq:Londonequation}
\end{multline}
where $\Pi(z)$ is the boxcar function, which is equal to 1 for $-d/2 \leq z \leq d/2$ and 0 otherwise, $\vec{r}=(x,y)$ and $\vec{R}$ the vortex positions.
The solution is obtained by decomposing $B_z(x,y,z)$ into its Fourier components in the $x-y$
plane
\begin{equation}
B_z(x,y,z)= \sum_{\vec{k}} b_z({\vec{k}},z)e^{-i\vec{k}\cdot\vec{r}}
 \label{eq:Fourier}
\end{equation}
where $\vec{k}$ is the reciprocal lattice vector of the flux lattice with
$k=|\vec{k}|=\sqrt{\frac{16\pi^2(m^2-mn+n^2)}{3D^2}}$, $m$,$n$ integer.
After matching the field and its derivative at the layer boundaries, we
determine the Fourier coefficients $b_z({\vec{k}},z)$ so that solutions are obtained inside and outside the single-layer FeSe. The width of the field
distribution at $z$ is then given by $\Delta B^2_z(z)= \langle B^2_z(z)\rangle-\langle B_z(z)\rangle^2=\sum
\limits_{\substack{k\neq 0}} b_{z}(k,z)^2$.
Averaging is over the $x$ and $y$ plane coordinates.
For a comparison with the measured broadening, $\Delta B_{z}(z)$ has to be weighted with the
normalized muon stopping distribution $n(z,E)$ so that
$\sigma_{\rm sc}^2(E)=\gamma_{\mu}^2 \int_{-\infty}^{\infty} \Delta
B^2_{z}(z) n(z,E) dz$.
 In contrast to the 3D case where $\sigma_{\rm sc} \propto \frac{1}{\lambda^2}$, in our 2D situation we find
that the field broadening is governed by the Pearl length scale
$\Lambda_P \equiv 2\lambda^2/d$ as expected for the vortex state in superconducting films with $d << \lambda$ ~\cite{Pearl}. For instance, taking into account that the superconducting layer is very thin and that
the dominating contribution to the observed field broadening comes from the muons stopping outside the layer ($d/2 \leq z \leq -d/2$), one finds that the Fourier coefficients can be expressed as
$b_{z}(k,z)\cong \frac{B_{appl}}{\Lambda_P} \frac{e^{-i k |z|}}{k}$.

\subsection{Determination of microscopic superconducting properties}

The Pearl length scale is directly related to the sheet superconducting carrier density
$n_s^{2D}= \frac{2 m_e^{*}}{\mu_0 e^2 \Lambda_P}$.
\begin{figure}[h]
\begin{center}
\includegraphics[width=1.0\columnwidth]{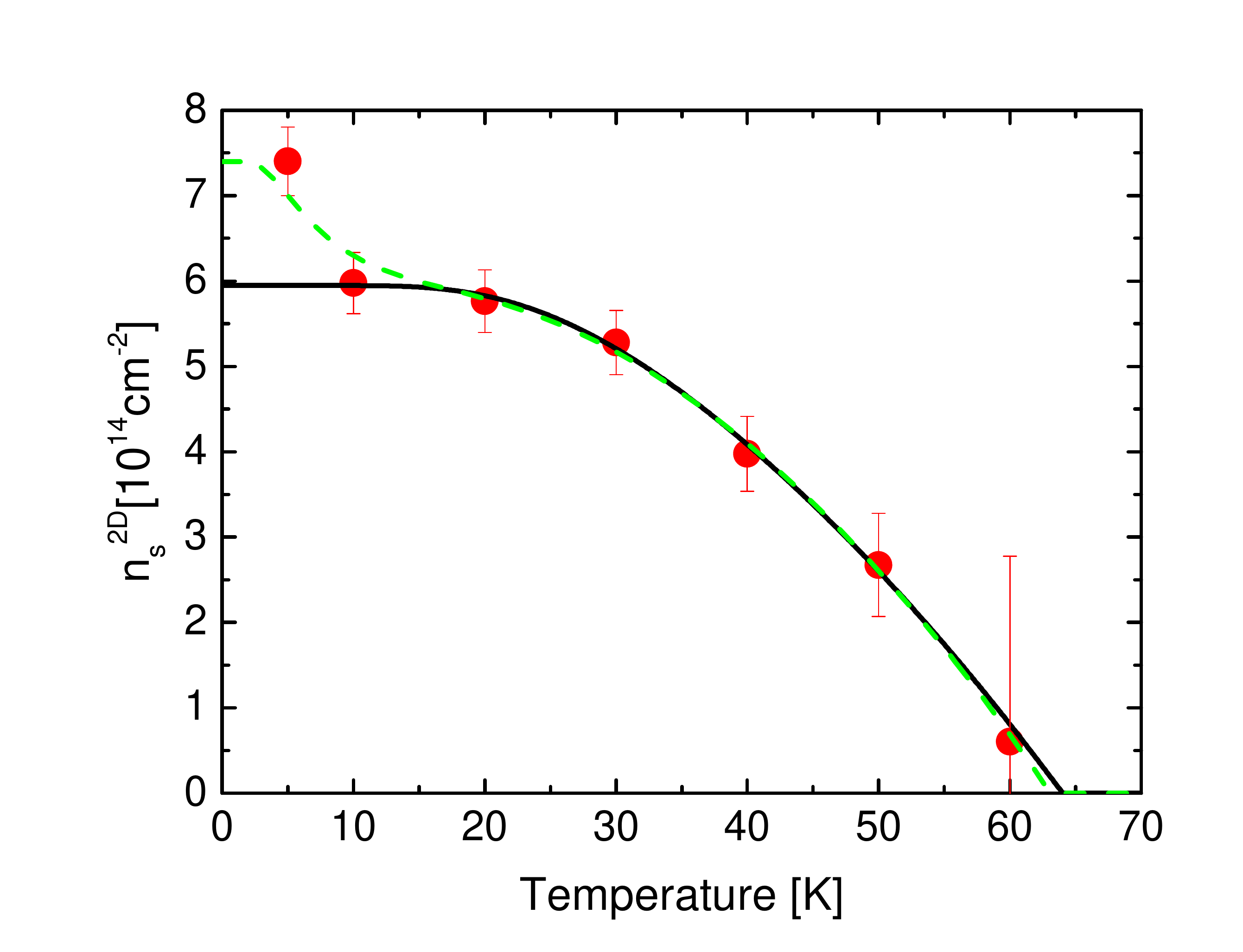}
\caption{\label{sheet density} \textbf{Temperature dependence of the superfluid sheet density.} Superfluid 2D density versus temperature for ultrathin
  FeSe. The solid curve is a fit with a BCS $s$-wave gap. For comparison a model assuming an additional small gap manifesting itself at low temperatures is shown as a dashed line.}
\end{center}
\end{figure}
Figure~\ref{sheet density} shows the temperature dependence of the sheet superfluid density in the ultrathin FeSe layer. Remarkably, $n_s^{2D}$ does not show any signs of phase
fluctations, which may be expected in a 2D-like superconductor, probably because of the strong coupling to the STO substrate \cite{Lee}. This temperature dependence can be well
fitted down to 10~K using a single-gap BCS $s$-wave model  (solid
line in Figure ~\ref{sheet density}).
The fit gives a gap value at zero temperature $\Delta(0)=10.2(1.1)$ meV and $T_c = 62(2)$ K. This gap value is consistent with several ARPES and STM measurements that find
values in the range 10-15 meV ~\cite{Lee,Liu,He,Tan,Huang}. Some STM measurements have reported gap structures with double peaks at $\approx$ 10 meV and 15-20 meV \cite{Wang2}. These differences may be due to differences in annealing conditions of the sample, protection layer or substrate preparation \cite{Zhang2014}.
It is worth noting that, since the muons uniformly probe the entire area of the sample, the measured parameters are sample average values, which may explain why our gap value is on the lower side of ARPES and STM values.
$\mu$SR is able to discriminate between different electronic phases. Our TF-$\mu$SR data can be fitted with a single superconducting component. Therefore, the results show that homogeneous superconductivity exists across the entire
FeSe/STO interface of size $\sim$ cm$^2$. By homogeneity, we mean here (a) homogeneous superconductivity on a scale of the order of the Pearl length scale or larger and (b) that on this scale there is no phase separation, e.g. in superconducting and non-superconducting regions. This does not exclude, however, local inhomogeneity at nano- or subnanoscale.
The gap to $T_c$ ratio $\frac{\Delta(0)}{k_B T_c} = 1.9(2)$ puts the single-layer FeSe in the category of the moderately strong-coupling superconductors.

Fitting the measured energy dependence of $\sigma_{\rm sc}(E)$ (Figure ~\ref{sigma_energy}) with our model
we obtain $\Lambda_P = 2.49(5) \times 10^4$ nm at 10 K.  From this number we estimate the density of paired
electrons to be $n_s^{2D} \simeq 6
\times 10^{14}$ cm$^{-2}$ (with the effective mass
$m^{*}=2.7 m_e$ \cite{Liu}). The choice of $m^{*}=2.7 m_e$ is confirmed by recent measurement of electron doped FeSe \cite{Wen2016},
where the effective mass of the electron band at M (relevant band in single layer FeSe) was found to lie between 2.7 $m_e$ and 3.5 $m_e$ (for a doping of 0.1-0.12 electron per Fe, which corresponds to the electron excess value in single-layer FeSe). Since $n_s^{2D} \propto \frac{m^{*}}{\Lambda_P}$, a value of 2.7 $m_e$ sets rather a lower limit for $n_s^{2D}$.
On the other hand disorder of the vortex lattice would increase the value of $\Lambda_P$. However, disorder contributes only quadratically to the measured spin relaxation rate so that even a contribution equal to the broadening associated with the vortex field would decrease $n_s^{2D}$ by $\sim$ 40 \%. Overall we assign an error to our estimate of the sheet carrier density of $\pm 30\%$.

Not many methods are able to determine the superconducting carrier density of very thin layers. More important, we would like to stress here that unlike other techniques our measurement provides a direct estimate of the paired carriers in
a buried FeSe layer.
The determination of this quantity is of relevance for instance to clarify its link to the enhancement of superconductivity with respect to the bulk counterpart and to understand the mechanism working at the interface between the FeSe layer and the substrate. Charge (electron) doping by ionic liquid gating \cite{Lei2016, Shiogai2017} or from deposited K atoms has been found important on its own to raise $T_c$. On the other hand, ARPES spectroscopy studies indicate that, beyond that, interface coupling may be necessary to get the highest $T_c$ close to liquid N$_2$ temperature \cite{Rebec2017}.
Assuming a dominant electron character, a Hall measurement (Hall coefficient $R_H=\frac{1}{n_e e} \simeq -3 \cdot 10^{-3} \rm{cm^{-3}}/C$) of thin ionic liquid gated FeSe flakes \cite{Lei2016} with $T_c \sim 48$ K  gives a carrier density at 50 K of $2 \times 10^{21} \rm{cm^{-3}}$. ARPES measurements of the electronic structure of single-layer FeSe with $T_c \approx 60$ K estimate an electron counting of $\sim 0.12$ electron/Fe \cite{Tan,Liu}, which corresponds to a similar volume density of carrier $2.2 \times 10^{21} \rm{cm^{-3}}$ in a 0.6 nm thick monolayer.
By contrast, other transport experiments have reported much higher numbers of Hall carriers.
From $R_H \simeq -0.33 \cdot 10^{-3} \rm{cm^{-3}}/C$, a value $n_e \simeq 1.9 \times 10^{22} \rm{cm^{-3}}$ has been inferred for 2.9 nm thick FeSe on MgO at 50 K \cite{Shiogai2017}. Similar high values have been found for one monolayer FeSe/STO capped by FeTe/Si \cite{Zhang2014}, indicating that the above mentioned agreement of transport and ARPES carrier determination may be fortuitous.
However, multiband effects and different types of carrier make it difficult to determine the relevant carrier density from the $R_H$ value, which is strongly  dependent on temperature and growth/annealing conditions \cite{Zhang2014} and may not give a reliable measure of the actual number of carriers that condense in the superconducting state.

A related question is the spatial extent of superconductivity in FeSe layers more than one monolayer thick. Although spectroscopic data indicate that additional layers have weak interlayer coupling with the second monolayer displaying semiconducting characteristics and charge neutrality \cite{Tan}, the question of the contribution of additional layers to the superconductivity of 1 ML FeSe is not fully understood.
Shiogai et al. \cite{Shiogai2017,Shiogai} used an electric double-layer transistor configuration, which allows at the same time electrostatic carrier doping and electrochemical thickness tuning, to identify a unified trend of $T_c$ vs $R_H$ for ultrathin FeSe layers on oxide substrates such as SrTiO$_3$, MgO and KTaO$_3$ and determine various length scales and critical thicknesses. Particularly, Hall measurements as a function of thickness allowed to determine the length scale of the charge distribution due to charge transfer from the substrate, $d_{CT}$, and the penetration length of the superconducting order parameter $\xi_{N}^{CT}$ in the layer above due to the proximity effect. For FeSe/STO $d_{\rm{CT}} \cong 4$ nm and $\xi_{\rm{N}}^{\rm{CT}} \cong 3.5$ nm, implying that ultrathin FeSe may exhibit high-Tc superconductivity on an effective length higher than that inferred by ARPES measurements of the electronic structure of $\geq$ 2 ML FeSe. Even allowing for band bending effects increasing the thickness of the charge transfer layer in the specific electric dipole layer configuration of Ref. [\onlinecite{Shiogai2017}], it appears reasonable to consider that proximity effects cannot be ignored in $>$ 1 ML thick FeSe layer. In this respect it is interesting to note that the
value $n_s^{2D} \simeq 1.4 \times 10^{14}$ cm$^{-2}$ obtained from the excess electron determination by ARPES is about a factor of four lower than the present determination of the superconducting carrier density $n_s^{2D} \simeq 6 \pm 2 \times 10^{14}$ cm$^{-2}$ of our heterostructure containing 1+4 FeSe layers.


%


The temperature dependence of the superfluid density (Figure ~\ref{sheet density}) may suggest an increase of this quantity at the lowest measured temperature, 5 K.
Since this effect appears only in a single data point we can only speculate about its significance. It might point to the presence of a second (small) gap effectively opening below 10 K.
We tried a two gap $s$+$s$ wave model to account for this low temperature increase. 
For this we analyzed our data with a phenomenological model by assuming two independent contributions to the total superfluid density but with a common $T_c$.
The functional form of the two gap model, which includes as a special case the single gap model, previously discussed, is ~\cite{Padamsee}:

\begin{equation}
\label{two_gap}
\frac{n_s^{2D}\left(T\right)}{n_s^{2D}\left(0\right)}=\omega\frac{n_s^{2D}\left(T, \Delta_{1}(0)\right)}{n_s^{2D}\left(0,\Delta_{1}(0)\right)}+(1-\omega)\frac{n_s^{2D}\left(T, \Delta_{2}(0)\right)}{n_s^{2D}\left(0,\Delta_{2}(0)\right)},
\end{equation}
where $\lambda\left(0\right)$ is the value of the penetration depth at $T=0$~K, $\Delta_{\rm i}(0)$ is the value of the $i$-th ($i=1$ or 2) superconducting gap at $T=0$~K and $\omega$ is the weighting factor of the band with the largest gap.

Each component of equation~\ref{two_gap} can be calculated within the local London approximation ($\lambda \gg \xi$)~\cite{Tinkham,Prozorov} as

\begin{equation}
\frac{n_s^{2D}\left(T, \Delta_{\rm i}(0)\right)}{n_s^{2D}\left(0, \Delta_{\rm i}(0)\right)}=1+2\int^{\infty}_{\Delta_{\rm i}(0)}\left(\frac{\partial f}{\partial E}\right)\frac{ EdE}{\sqrt{E^2-\Delta_{\rm i}\left(T\right)^2}},
\end{equation}
where $f=\left[1+\exp\left(E/k_{\rm B}T\right)\right]^{-1}$ is the Fermi function, and $\Delta_{\rm i}\left(T\right)=\Delta_{\rm i}(0)\delta\left(T/T_{\rm c}\right)$. The temperature dependence of the gap is parametrized by the expression $\delta\left(T/T_{\rm c}\right)=\tanh\left\{1.82\left[1.018\left(T_{\rm c}/T-1\right)\right]^{0.51}\right\}$, which well represents the temperature dependence of a BCS gap ~\cite{Carrington}.

A fit is shown as dashed line in Figure \ref{sheet density} yielding for the main gap $\Delta(0)=10.5(1.6)$ meV (in agreement with the single-gap fit)  and the putative small gap $\Delta(0)=1.3(6)$ meV with relative weight 0.23(4).
Another possibility may be some proximity contribution of the additional 4 monolayers of FeSe modifying the gap structure.
Further measurements are needed to elucidate this point, as well as the question about the possible presence of additional small gaps at much lower temperature and their nodal structure.

To conclude, by measuring \textit{ex situ} the depth and temperature
dependence of the local field distribution in a heterostructure
containing a buried superconducting ultrathin FeSe layer,
we detect the formation of a vortex state below $T_c \cong 60$ K and quantify the superfluid density of 1+4 ML FeSe. The temperature dependence can be well explained by a single BCS $s$-wave gap of 10.2(1.1) meV. The $\mu$SR spectra show that the vortex state
and superconductivity are homogeneously formed across the entire interface over a sample with a sizeable amount of charges condensing below $T_c \approx 62$ K. This shows that superconductivity in the buried interface has stable character and that inhomogeneities or imperfections of the substrate or of the overlayers do not hamper the formation of a superconducting state nor sizeably modify its properties.
A very sensitive magnetic probe such as polarized muons do not see indication of static or dynamic magnetism.
The simple structure of single-layer FeSe, its high $T_c$ with $s$-wave type of gap and rather clean BCS character make it an ideal system to develop a microscopic understanding of high-$T_c$ superconductivity.

\textbf{Acknowledgments}
The $\mu$SR experiments were performed at the Swiss Muon Source, Paul Scherrer Institut, Villigen, Switzerland. We thank A. Suter for fruitful discussions.


\begin{thebibliography}{99}

\bibitem{Bednorz}J. G. Bednorz  \& K. A. M\"uller  Possible high-T$_{\rm c}$ superconductivity in Ba-La-Cu-O system, \textit{Z. Phys. D.} \textbf{64}, 189-193 (1986).

\bibitem{Schilling} A. Schilling, M. Cantoni, J. D. Guo,  \& H. R. Ott, Superconductivity above 130 K in the Hg-Ba-Ca-Cu-O system. \textit{Nature} \textbf{363}, 56-58 (1993).

\bibitem{Hosono}  Y. Kamihara, T, Watanabe, M. Hirano, \& H. Hosono, Iron-Based Layered Superconductor La[O$_{1-x}$F$_x$]FeAs $(x = 0.05-0.12)$ with $T_c = 26$ K. \textit{J. Am. Chem. Soc.} \textbf{130}, 3296 (2008).

\bibitem{Hsu} F.C. Hsu, \textit{et al.} Superconductivity in the PbO-type structure $\alpha$-FeSe. \textit{Proc. Natl. Acad. Sci. U.S.A.} \textbf{105}, 14262-14264 (2008).

\bibitem{Johnston} D.C. Johnston, The puzzle of high temperature superconductivity in layered iron pnictides and chalcogenides. \textit{Adv. Phys.} {\bf 59}, 803-1061 (2010).

\bibitem{Paglione} J. Paglione, R. L. \& Greene, High-temperature superconductivity in iron-based materials. \textit{Nat. Phys.} {\bf 6}, 645-658 (2010).

\bibitem{Stewart} G. R. Stewart, Superconductivity in iron compounds. \textit{Rev. Mod. Phys.} {\bf 83}, 1589-1652 (2011).
%
\bibitem{Wang} F. Wang, \& D. H. Lee, The electron-pairing mechanism of iron-based superconductors. \textit{Science} {\bf 332}, 200-204 (2011).
%
%
\bibitem{Wang2} Q. Y. Wang, \textit{et al.} Interface-Induced High-Temperature Superconductivity in Single Unit-Cell FeSe Films on SrTiO$_3$. \textit{Chin. Phys. Lett.} \textbf{29}, 037402 (2012).

\bibitem{Lee} J. J. Lee, \textit{et al.} Interfacial mode coupling as the origin of the enhancement of $T_c$ in FeSe films on SrTiO$_3$. \textit{Nature} \textbf{515}, 245 (2014).


\bibitem{Liu} D. F. Liu, \textit{et al.} Electronic origin of high-temperature superconductivity in single-layer FeSe superconductor. \textit{Nat. Commun.} \textbf{3}, 931 (2012).

\bibitem{He} S. L. He, \textit{et al.}  Phase diagram and electronic indication of high-temperature superconductivity at 65 K in single-layer FeSe films. \textit{Nat. Mater.} \textbf{12}, 605-610 (2013).

\bibitem{Tan} S. Y. Tan, \textit{et al.} Interface-induced superconductivity and strain-dependent spin density waves in FeSe/SrTiO$_3$ thin films. \textit{Nat. Mater.} \textbf{12}, 634-640 (2013).

\bibitem{Rebec2017} S. N. Rebec \textit{et al.} Coexistence of Replica Bands and Superconductivity in FeSe Monolayer Films, S. N. Rebec, T. Jia, C. Zhang, M. Hashimoto, D.-H. Lu, R. G. Moore, and Z.-X. Shen,
Phys. Rev. Lett. {\bf 118}, 067002 (2017).



\bibitem{Wang2017} Z. Wang, C. Liu, Y. Liu and J. Wang High-temperature superconductivity in one-unit-cell FeSe films. \textit{J. Phys.: Condens. Matter} \textbf{29}, 153001 (2017).

\bibitem{Shiogai} J. Shiogai, \textit{et al.} Electric-field-induced superconductivity in electrochemically etched ultrathin FeSe films on SrTiO$_3$ and MgO. \textit{Nat. Phys.} \textbf{12}, 42 (2016). doi:10.1038/nphys3530

\bibitem{Lei2016}B. Lei  \textit{et al.}, Evolution of High-Temperature Superconductivity from a Low-$T_c$ Phase Tuned by Carrier Concentration in FeSe Thin Flakes.  \textit{Phys. Rev. Lett.} \textbf{116},077002 (2016).

\bibitem{Zhang2} W. Zhang, \textit{et al.} Direct observation of high-temperature superconductivity in one-unit-cell FeSe films. \textit{Chin. Phys. Lett.} \textbf{31}, 017401 (2014).

\bibitem{Ding2016} H. Ding \textit{et al.} , High-Temperature Superconductivity in Single-Unit-Cell FeSe Films on Anatase TiO$_2$(001) \textit{Phys. Rev. Lett.} \textbf{117}, 067001 (2016).

\bibitem{Wen2016} C.H.P. Wen,  \textit{et al.} Anomalous correlation effects and unique phase diagram of electron-doped FeSe revealed by photoemission spectroscopy. \textit{Nat. Commun.} \textbf{7}, 10840 (2016).

\bibitem{Ge} J. F. Ge, \textit{et al.} Superconductivity above 100 K in single-layer FeSe films on doped SrTiO$_3$. \textit{Nat. Mater.} \textbf{14}, 285-289 (2015).

\bibitem{Fan2015} Q. Fan, \textit{et al.} Plain \textit{s}-wave superconductivity in single-layer FeSe on SrTiO$_3$ probed by scanning tunneling microscopy. \textit{Nat. Phys.} \textbf{11}, 946 (2015).

\bibitem{Huang} D. Huang, \textit{et al.} Revealing the Empty-State Electronic Structure of Single-Unit-Cell FeSe/SrTiO$_3$. \textit{Phys. Rev. Lett.} \textbf{115}, 017002 (2015).

\bibitem{Zhang2014} W. Zhang \textit{et al.} Interface charge doping effects on superconductivity of single-unit-cell FeSe films on SrTiO$_{3}$ substrates. \textit{Phys. Rev. B} \textbf{89}, 060506(R) (2014).


\bibitem{Zhang2015} Z. Zhang, \textit{et al.} Onset of the Meissner effect at 65 K in FeSe thin film grown on Nb-doped SrTiO$_3$ substrate. \textit{Sci. Bulletin} \textbf{60 (14)}, 1301-1304 (2015).


\bibitem{Shiogai2017} J. Shiogai, T. Miyakawa, Y. Ito, T. Nojima, and A. Tsukazaki, Unified trend of superconducting transition temperature versus Hall coefficient for ultrathin FeSe
films prepared on different oxide substrates, Phys. Rev. B \textbf{95}, 115101 (2017).

\bibitem{Saadaoui} H. Saadaoui, Z. Salman, T. Prokscha, A. Suter, B.M. Wojek, E. Morenzoni, Zero-field Spin Depolarization of Low-Energy Muons in Ferromagnetic Nickel and Silver Metal
\textit{Physics Procedia} \textbf{30}, 164 (2012).

\bibitem{Morenzoni2004} E. Morenzoni, \textit{et al.} Nano-scale thin film investigations with slow polarized muons. \textit{J. Phys.: Condens. Matter} \textbf{16}, S4583 (2004).

\bibitem{Prokscha} T. Prokscha, \textit{et al.} The new mu E4 beam at PSI: A hybrid-type large acceptance channel for the generation of a high intensity surface-muon beam. \textit{Nucl. Instrum. Methods Phys. Res., Sect. A} \textbf{595}, 317 (2008).

\bibitem{Morenzoni} E. Morenzoni, \textit{et al.} Low-energy $\mu$SR at PSI: present and future. \textit{Physica B: Condensed Matter} \textbf{289}, 653 (2000).
%

\bibitem{Niedermayer} Ch. Niedermayer, \textit{et al.} Direct observation of a flux line lattice field distribution across an YBa$_2$Cu$_3$O$_{7-\delta}$ surface by low energy muons.
\textit{Phys. Rev. Lett.} \textbf{83}, 3932 (1999).
%
\bibitem{Morenzoni11} E. Morenzoni, \textit{et al.} The Meissner effect in a strongly underdoped cuprate above its critical temperature. \textit{Nat. Commun.} {\bf 2}, 272 (2011).
%
\bibitem{DiBernardo} A. Di Bernardo \textit{et al.} Intrinsic Paramagnetic Meissner Effect Due to $s$-Wave Odd-Frequency Superconductivity, \textit{Phys. Rev. X} \textbf{5}, 041021 (2015).
%
\bibitem{Al Ma’Mari} F. Al Ma\textsc{\char13}Mari \textit{et al.} Beating the Stoner criterion using molecular interfaces, \textit{Nature} \textbf{524}, 69 (2015).

\bibitem{Eckstein} W. Eckstein, \textit{Computer Simulations of Ion-Solid Interactions}, (Springer Verlag Berlin, Heidelberg and New York, 1991).
%
\bibitem{Morenzoni2002} E. Morenzoni \textit{et al.} Implantation studies of keV positive muons in thin metallic layers. \textit{Nucl. Instr. and Methods} {\bf 192}, 254-266 (2002).

\bibitem{Luke2} G. M. Luke, \textit{et al.} Time-reversal symmetry breaking superconductivity in Sr$_2$RuO$_4$. \textit{Nature} \textbf{394}, 558 (1998).
%
\bibitem{Biswas} P. K. Biswas,  \textit{et al.} Evidence for superconductivity with broken time-reversal symmetry in locally noncentrosymmetric SrPtAs. Phys. Rev. B \textbf{87}, 180503(R) (2013).
%
\bibitem{Kubo}  R. A. Kubo, stochastic theory of spin relaxation. \textit{Hyperfine~Interact.} \textbf{8}, 731 (1981).
%
\bibitem{Suter} A. Suter , B. M. Wojek, Musrfit: A Free Platform-Independent Framework for $\mu$SR Data Analysis, \textit{Physics Procedia.} \textbf{30}, 69 (2012).
%
\bibitem{Sonier} J. E. Sonier, J. H. Brewer,  R. F. \& Kiefl, $\mu$SR studies of the vortex state in type-II superconductors. \textit{Rev. Mod. Phys.} \textbf{72}, 769 (2000).
%
\bibitem{Clem} J. R. Clem, Two-dimensional vortices in a stack of thin superconducting films: A model for high-temperature superconducting multilayers. \textit{Phys. Rev. B} \textbf{43}, 7837 (1991).
%
\bibitem{Brandt2005a} E.H. Brandt, Ginzburg-Landau vortex lattice in superconductor films of finite thickness. \textit{Phys. Rev. B} \textbf{71},014521 (2005).
%
\bibitem{Carneiro2000} G. Carneiro,  E. H. \& Brandt,  Vortex lines in films: Fields and interactions. \textit{Phys. Rev. B} {\textbf 61}, 6370 (2000).
%
\bibitem{Brandt2005}  E.H. Brandt, Ginzburg-Landau vortex lattice in superconductor films of finite thickness. \textit{Phys. Rev. B} \textbf{71},014521 (2005).
%
\bibitem{Pearl} J. Pearl, Current distribution in superconducting films carrying quantized fluxoids. \textit{Appl. Phys. Lett.} \textbf{5}, 65 (1964).
%
\bibitem{Padamsee} H. Padamsee, J. E. \& Neighbor, C. A. \& Shiffman, Quasiparticle phenomenology for thermodynamics of strong-coupling superconductors. \textit{J.~Low~Temp.~Phys.} {\bf 12}, 387 (1973).
%
\bibitem{Tinkham} M. Tinkham, Introduction~to~Superconductivity. (McGraw-Hill, New~York, 1975).
%
\bibitem{Prozorov} R. Prozorov, R. W. \& Giannetta, Magnetic penetration depth in unconventional superconductors. \textit{Supercond.~Sci.~Technol.} \textbf{19}, R41 (2006).
\bibitem{Carrington} A. Carrington, F. \& Manzano, Magnetic penetration depth of MgB$_2$. \textit{Physica~C} \textbf{385}, 205 (2003).
%
%
%
%

\end{thebibliography}
\end{document}